\documentstyle[amstex,amssymb,righttag,12pt]{article}

\newtheorem{pro}{Proposition}

\def\adots{\mathinner{\mkern2mu\raise1pt\hbox{.}\mkern2mu
\raise4pt\hbox{.}\mkern2mu\raise7pt\hbox{.}\mkern1mu}}

\begin{document}

\title{\Large\sc Fully Supersymmetric Hierarchies From \\
A Energy Dependent Super Hill Operator}

%Energy Dependent Super Hill Operator \\
%and Related Integrable Hierarchies}

\author{Q. P. Liu\thanks{On leave of absence from
Beijing Graduate School, CUMT, Beijing 100083, China.}
\thanks{Supported by {\em Beca para estancias temporales
de doctores y tecn\'ologos extranjeros en
Espa\~na: SB95-A01722297}.}\\
 Departamento de F\'\i sica Te\'orica,\\ Universidad Complutense,\\ 
E28040-Madrid, Spain.}

\date{}

\maketitle

\begin{abstract}
A super Hill operator with energy dependent potentials
is proposed and the associated integrable hierarchy is 
constructed explicitly. It is shown that in the general
case, the resulted hierarchy is multi-Hamiltonian system.
The Miura type transformations and modified hierarchies
are also presented.
\end{abstract}

\newpage

\section{Introduction}
Schr\"odinger equation with energy dependent potential
is first studied by Jaulent and Miodek\cite{jm} and there
in the simplest case, the associated nonlinear
evolution equations are solved 
by means of Inverse Scattering Transformation.
The problem has been generalized to more general case
in\cite{luis}
and is further shown that the resulted flows are Bi-Hamiltonian
system.

The remarkable multi-Hamiltonian structures behind have been
 explored
and Miura type maps are obtained in a series papers of
Antonowicz and Fordy\cite{af1} - \cite{af4}. The Lie algebraic
reason for constructing Miura map is provided by Marshall
\cite{ian1}\cite{ian2} and this subsequently leads to some new 
results for the Ito's
system\cite{lm}. The most recent result for these hierarchies
is their relationship with the zero sets of the tau function
of the KdV hierarchy\cite{mmm}.

The generalizations of linear problems with 
energy potentials are interesting and
begins with the third order operator or Lax
operator for Boussinesq equation\cite{afl}. Unlike the
Schr\"odinger case, one does not have 
arbitrary polynomial dependent expansions
here and to have interesting results, one only obtains
four cases(see \cite{afl} for details). 
Similarly, Toda system is generalised this way\cite{liu1}.

We notice that integrable systems have super counterparts. 
Indeed there exist two different types of generalizations:
 fermionic or supersymmetric. For the celebrated KdV system, 
this corresponds to Kupershmidt's super KdV\cite{bk} 
and Manin-Radul's
super KdV\cite{mr}, respectively. While both generalizations
are interesting from the mathematical viewpoint, 
it is believed that the supersymmetric extensions
are physically relevant. In \cite{af4}, Kupershmidt's spectral
problem for super KdV
is generalized to the context of energy dependent potentials.

The aim of the present paper is to present  fully
supersymmetric integrable systems resulted
from a linear super operator with energy dependent
potentials. We will show that like Schr\"odinger
operator case,
the resulted systems are multi-Hamiltonian nature and
have multi-step modifications. Thus, the remarkable
algebraic structures revealed in \cite{af1}-\cite{af4}
are retained for our new supersymmetric systems. The 
simplest example in this construction
includes one of $N=2$ supersymmetric
KdV system\cite{lam}.

The paper is organized as follows. In section two, we 
propose the linear problem and construct the related 
isospectral flows. We also construct the matrix operators
which are our candidates for Hamiltonian operators.
In section three, we proceed to construct the Miura type
maps which serve as a simple way to prove
some claims made in section two.
Section four contains some interesting examples.

\section{Linear Problem}
We start with the following super linear operator
\begin{equation}
L=\varepsilon D^3+uD+\alpha,
\end{equation}
where $D=\partial_{\vartheta}+\vartheta\partial$ $\,$ is the super
derivation with $\vartheta$ a Grassman odd variable and
$\partial={\partial \over\partial x}$;
$\varepsilon=\varepsilon(\lambda)$ is a bosonic
parameter depending on  the spectral parameter $\lambda$; 
$u=u(\lambda;\vartheta,x,t)$ is bosonic variable and
$\alpha=\alpha(\lambda;\vartheta,x,t)$ is fermionic variable.

To obtain isospectral flows associated to $L$,
we consider the linear problem $L\psi=0$ together
with the time evolution of wave function:
\begin{equation}
\psi_t=P\psi, \qquad P=b\partial+\beta D+c,
\end{equation}
then by simple calculation, we have 
\begin{align*}
 L_t -[P,L]=&
u_t D+\alpha_t
-\varepsilon \big(2\beta -(Db)\big)\partial^2+
\varepsilon \big(b_x +(D\beta)\big)D^3 \\
&+\big(\varepsilon (Db)_x-\varepsilon \beta_x+
\varepsilon (Dc)+u(Db)-2u\beta\big)D^2\\
&-
\big(bu_x+\beta (cu)-\varepsilon (D\beta)_x-\varepsilon c_x
-u(D\beta)\big)D\\
&-\big(b\alpha_x+\beta (D\alpha)-\varepsilon (Dc)_x-u(Dc)\big),
\end{align*}
it is easy to see that the usual Lax equation
$L_t=[P, L]$ will not lead to any consistent equation. To
have meaningful results,  we introduce 
\[
Q=\big((Db)-2\beta\big)D+b_x+(D\beta),
\]
and consider
\begin{align*}
[P,L]+QL=&\big(-\varepsilon (Db)_x+\varepsilon \beta_x-\varepsilon (Dc)\big)D^2
+\big(bu_x+\beta (Du)-\varepsilon (D\beta)_x\\
&-\varepsilon c_x-u(D\beta)+(Db)(Du)-2\beta (Du)-(Db)\alpha
+2\beta\alpha\\
&+b_xu+(D\beta) u\big)D+b\alpha_x+\beta (D\alpha)-\varepsilon (Dc)_x
-u(Dc)\\
&+(Db)(D\alpha)-2\beta (D\alpha)+
b_x\alpha+(D\beta)\alpha,
\end{align*}
thus we have to choose $c=-b_x+(D\beta)$ and then
\[
L_t=[P,L]+QL,
\]
gives us 
\begin{align*}
u_t&=(bu)_x-\beta (Du)-2\varepsilon (D\beta)_x +\varepsilon 
b_{xx}+(Db)(Du)-(Db)\alpha+2\beta
\alpha,\\
\alpha_t&=(b\alpha)_x+D(\beta\alpha)+\varepsilon 
(Db)_{xx}-\varepsilon\beta_{xx}+
u(Db)_x-u\beta_x+(Db)(D\alpha),
\end{align*}
which can be written neatly as
\begin{equation}\label{basic}
{\boldsymbol u}_t=J{\boldsymbol R},
\end{equation}
where
\[
{\boldsymbol u}=\left(\begin{array}{c}u\\ \alpha \end{array}\right),
\qquad {\boldsymbol R}=\left(\begin{array}{c}(Db)-\beta\\
 b\end{array}\right),
\]
\begin{equation}\label{J}
J=\left(\begin{array}{cc}2\varepsilon D^3+2\alpha -(Du)&
-\varepsilon\partial^2 -\alpha D+\partial u\\
\varepsilon\partial^2+u\partial+D\alpha &\alpha\partial+\partial\alpha
\end{array}\right). 
\end{equation}

To obtain the evolution equations, we now specify
the $\varepsilon$, $u$ and $\alpha$ in the following way
\begin{equation}\label{spec}
\varepsilon =\sum_{i=0}^{n}\varepsilon_i\lambda^i,\qquad
u=\sum_{i=0}^{n} u_i\lambda^i, \qquad
\alpha=\sum_{i=0}^{n}\alpha_i\lambda^i,
\end{equation}
with the above choice(\ref{spec}), the equation (\ref{basic})
is in the form
\begin{equation}\label{basic0}
\sum_{i=0}^{n}\lambda^i \boldsymbol{u}_{it}=
\left(\sum_{i=0}^{n}J_i \lambda^i \right)\boldsymbol{R},
\end{equation}
where 
$
\boldsymbol{u}_i=(u_i,\alpha_i)^T
$
and
\begin{equation}\label{basic1}
J_i=\left(\begin{array}{cc}2\varepsilon_i D^3+2\alpha_i -(Du_i)&
-\varepsilon_i\partial^2 -\alpha_i D+\partial u_i\\
\varepsilon_i\partial^2+u_i\partial+D\alpha_i &\alpha_i\partial+\partial\alpha_i
\end{array}\right). 
\end{equation}

We assume that the $\boldsymbol{R}$ has following expansion
with respect to the spectral parameter $\lambda$
\[
\boldsymbol{R}=\sum_{i=0}^{m}\boldsymbol{R}_{m-i}\lambda^i,
\]
then the coefficients of different powers of $\lambda$
of the equation(\ref{basic0}) give us
%\begin{subequations}
\begin{equation}\label{re1}
\begin{aligned}
\boldsymbol{u}_{0t}&=J_0\boldsymbol{R}_m,\\
\boldsymbol{u}_{1t}&=J_0\boldsymbol{R}_{m-1}+J_1\boldsymbol{R}_m,\\
\vdots\\
\boldsymbol{u}_{nt}&=J_0\boldsymbol{R}_{m-n}+J_1\boldsymbol{R}_{m-n+1}+
\hdots+J_n\boldsymbol{R}_m,
\end{aligned}
\end{equation}
%\eqno(\mbox{2.9a})
\begin{equation}\label{re2}
\begin{aligned}
J_0\boldsymbol{R}_{i-n}+J_1\boldsymbol{R}_{i-n+1}+\hdots +
J_n\boldsymbol{R}_i=0, \qquad i=0,\hdots, m-1.
\end{aligned}
\end{equation}
%\end{subequations}
%\eqno(\mbox{2.9b})
%\setcounter{equation}{9}
From the above systems(\ref{re1}-\ref{re2}),
 we see that $\boldsymbol{R}_m$ is 
not determined and we have two basic cases:
\begin{itemize}
\item $u_n=-1$, $\alpha_n=0$.

In this case, the last equation of (\ref{re1}) takes the same
form of (\ref{re2}) with $i=m$, which enables us to determine
$\boldsymbol{R}_m$ in principal.  This case is refereed as KdV case.
\item $u_0=$constant, $\alpha_0=$ (fermionic) constant

This case leads to $\boldsymbol{R}_m=0$ for compatibility and is refereed as
Harry Dym case.
\end{itemize}

Since the second case can be studied similarly, 
next we will only consider the first case in detail.

The evolution equation(\ref{re1}) can be reformed as
\begin{equation}\label{evo}
\left(\begin{array}{c}\boldsymbol{u}_0\\
\vdots\\ \boldsymbol{u}_{n-1}\end{array}\right)_{t_{m}}
=\left(\begin{array}{ccc} 0&& J_0\\
&\adots&\vdots\\ J_0&\hdots&J_{n-1}\end{array}\right)
\left(\begin{array}{c}\boldsymbol{R}_{m-n+1}\\
\vdots\\ \boldsymbol{R}_{m}\end{array}\right),
\end{equation}
and the recursion relation (\ref{re2}) can be written as
\begin{equation}
B_i {\boldsymbol R}^{(k)}=B_{i-1} {\boldsymbol R}^{(k+1)}, 
\qquad i=1,\hdots, n,
\end{equation}
where 
\begin{equation}\label{ha}
B_i=\left(\begin{array}{cc}
\begin{array}{ccc}
0&&J_0\\ &\adots&\vdots \\ J_0&\hdots &J_{i-1}
\end{array}
&0\\
0& \begin{array}{ccc}-J_{i+1}&\hdots&-J_n\\
\vdots&\adots&\\
-J_n&&0\end{array}\end{array}\right),
\end{equation}
and
\[
{\boldsymbol R}^{(k)}=({\boldsymbol R}_{k-n+1},\hdots,{\boldsymbol R}_{k})^T,
\]
these operators $B_i$ are our candidates of Hamiltonian
operators. In order to obtain Hamiltonian description
of the evolution system(\ref{evo}), we need to prove
\begin{itemize}
\item[(i)] $B_i$ are Hamiltonian operators;
\item[(ii)] $J{\cal{R}}=0$ admits the formal power series solution
${\cal{R}}=\sum_{i=0}^{\infty} \boldsymbol{R}_i\lambda^{-i}$;
\item[(iii)] ${\boldsymbol R}^{(i)}$ can be written as variational derivatives
of some functionals ${\cal H}_i$.
\end{itemize}

With the assumption that above statements are proved,
we now has
\begin{equation}
\boldsymbol{U}_{t_{m}}=B_{n-k}\delta {\cal H}_{m+k},
\qquad k=0,\hdots,n,
\end{equation}
where ${\boldsymbol U}=({\boldsymbol u}_0,\hdots,
{\boldsymbol u}_{n-1})^T$ and $\delta$ denotes the 
variational derivative with respect to ${\boldsymbol U}$. 
Thus, our system 
is a $(n+1)$ Hamiltonian system.

Now we prove the statement (ii-iii). To this end, we introduce
$\eta=D(\ln \psi)$, then $L\psi=0$ becomes
\begin{equation} \label{ricati}
\varepsilon (\eta_x+\eta (D\eta))+u\eta+\alpha=0,
\end{equation}
it is easy to see that (\ref{ricati}) has the following
solution
$\eta=\sum^{s}_{-\infty}\eta_{-j}\lambda^{j}$
for certain $s$. It is also ready to see that each $\eta_j$
provides us a conserved quantity in principal. Next 
we show that the solution of (\ref{ricati}) will supply a set solution
for $J{\cal{R}}=0$. For clarity, we formulate it as
\begin{pro}
For each solution $\eta $ of the equation (\ref{ricati}),
its variational derivative, with respect to
$(u,\alpha)$, provides us a solution for $J{\cal R}=0$.
\end{pro}
{\em Proof}: We introduce an additional variable
$y$ so that the equation (\ref{ricati}) is written as
\begin{equation}\label{map}
\begin{aligned}
u&=\varepsilon\big((D\eta)-y\big),\\
\alpha&=\varepsilon\big(\eta_x-2\eta(D\eta)+\eta y\big),
\end{aligned}
\end{equation}
notice that the equation(\ref{map}) serves as a map between $(u,\alpha)$ and 
$(y,\eta)$. Thus, we have the following formula
\begin{equation}\label{var}
\left(\begin{array}{c}\delta_y\\ \delta_{\eta}
\end{array}\right)=F^{\dag}\left(\begin{array}{c}
\delta_u\\ \delta_{\alpha}\end{array}\right),
\end{equation}
where $\delta_v={\delta\over\delta v}$ and 
\[
F=\varepsilon\left(\begin{array}{cc}-1&D\\
\eta&-\partial-2\eta D-2(D\eta)+y\end{array}\right),
\]
is the Fr\'echet derivative of (\ref{map})
and $\dag$ denotes adjoint.

Acting the equation (\ref{var}) on $\eta$ and denoting
$\xi=\delta_u \eta$, $p=\delta_{\alpha} \eta$, we obtain
\begin{equation}\label{var1}
\begin{aligned}
\xi-\eta p&=0,\\
\varepsilon\big[(D\xi)+p_x+2\eta (Dp)-4(D\eta)p+yp\big]&=1.
\end{aligned}
\end{equation}
We claim that the solution $(\xi, p)$ of the 
system(\ref{var1}) 
provides us a solution of $J{\cal R}=0$. To see the validity of
this claim, we eliminate the variable $\xi$ in (\ref{var1})
and have
\begin{equation}
\varepsilon\big[p_x+\eta(Dp)+yp-3(D\eta)p\big]=1.
\end{equation}
Differentiating above equation leads to
\begin{equation}\label{var2}
\varepsilon\big[(Dp)_x+(y-2(D\eta))(Dp)+(Dy-3\eta_x+\eta y-
3\eta(D\eta)p\big]=\eta,
\end{equation}
\begin{equation}\label{var3}
\begin{aligned}
\varepsilon \big[p_{xx}+&(\eta_x-2\eta y+5\eta (D\eta))(Dp)+
(y_x-3(D\eta_x)-\eta(Dy)\\
+&3\eta \eta_x-y^2+6y(D\eta)-9(D\eta)^2)p\big]
 +y-3(D\eta)=0,
\end{aligned}
\end{equation}
now using above formulae(\ref{var2})-(\ref{var3}) 
and meanwhile keeping
in mind the mapping(\ref{map}), one can easily show that
\[
2\varepsilon(D\xi_x)+2\alpha\xi-(Du)\xi-\varepsilon p_{xx}-\alpha(Dp)
+(up)_x=0.
\]

Similarly, we can check the 
\[
\varepsilon\xi_{xx}+u\xi_x+D(\alpha \xi)+2\alpha p_x+\alpha_x
p=0,
\]
is an identity. Last two equation is nothing but $J{\cal R}=0$ and
the proposition is proved.$\Box$

{\em Remark}: Solvability is justified by supplying a set of solutions
as above. So the strategy used here is different from bosonic case,
where one is able to prove this fact directly(cf. \cite{af3}).

\section{Miura Maps and Modifications}
To construct the Miura type map for the systems presented
in last section, we first consider the basic case: 
$u\to u-\lambda$, $\alpha\to\alpha$.

By the following factorization
\[
L=(D+\theta_1)(D+\theta_1+\theta_2)(D+\theta_2),
\]
we have
\begin{equation}\label{basicmiura}
\begin{aligned}
u&=w_x+(D\theta)+\theta(Dw), \\
\alpha&=(Dw)_x+(D\theta)(Dw),
\end{aligned}
\end{equation}
where we made redefinitions of coordinates
$\theta_1=\theta$ and $\theta_2=Dw$ for convenience.

The Fr\'echet derivation of the map (\ref{basicmiura})
and its adjoint are
\begin{align*}
m&=\left(\begin{array}{cc}
\partial +\theta D&D-(Dw)\\
D\partial+(D\theta)D&(Dw)D\end{array}\right),
\\ 
m^{\dag}&=\left(\begin{array}{cc}
-\partial +D\theta &D\partial-D(D\theta)\\
D+(Dw)&D(Dw)\end{array}\right),
\end{align*}
and we can verify the following identity holds
\[
mKm^{\dag}=J,
\]
where $K=\left(\begin{array}{cc}0&1\\ -1&0\end{array}\right)$
and $J$ is given by (\ref{J}).

Modifying the map (\ref{basicmiura}) with a parameter
$\gamma$ so that we have
\begin{equation}
u=\gamma w_x+(D\theta)+\theta(Dw),\qquad
\alpha=\gamma(Dw)_x+(D\theta)(Dw).
\end{equation}

With the preparation, we now follow the method
presented in \cite{af3} and construct
the Miura maps. Since the construction following closely from
one presented in \cite{af3}-\cite{af4}, we just present final
results here. Factorizing $J=\sum_{i=0}^{n}\lambda^i J_i$
in the following way
\begin{equation}\label{fac}
J=(m_0,m_1,\hdots,m_n)K\Lambda_k (m_{0}^{\dag},m_{1}^{\dag},
\hdots, m_{n}^{\dag})^T,
\end{equation}
where
\begin{align*}
m_i&=\left(\begin{array}{cc}
\gamma_i\partial +\theta_i D&D-(Dw_i)\\
\gamma_iD\partial+(D\theta_i)D&(Dw_i)D\end{array}\right),
\\
m^{\dag}_{i}&=\left(\begin{array}{cc}
-\gamma_i\partial +D\theta_i &\gamma_i D\partial-D(D\theta_i)\\
D+(Dw_i)&D(Dw_i)\end{array}\right),
\end{align*}
and
\[
\Lambda_k=\left(\begin{array}{cc}
\begin{array}{ccc}1&\hdots&\lambda^{k-1}\\
\vdots&\adots&\\ \lambda^{k-1}&&\end{array}&0\\
0&\begin{array}{ccc}0&&\lambda^k\\
&\adots&\vdots\\
\lambda^k&\hdots&\lambda^n\end{array}\end{array}\right),
\]
and comparing the coefficients of $\lambda$ of 
the equation(\ref{fac}) we have 
\begin{equation}\label{par}
\begin{aligned}
\varepsilon_k&=\sum^{k}_{i=0}\gamma_i,\qquad k=0,\hdots,r-1,\\
\varepsilon_k&=\sum^{n}_{i=k}\gamma_i, \qquad k=r,\hdots,n,
\end{aligned}
\end{equation}
\begin{equation}\label{m1}
%\begin{aligned}
u_k={1\over 2}\sum_{i=0}^{k}{\cal{W}}_{i.k-i},\qquad
\alpha_k={1\over 2}\sum_{i=0}^{k}\Omega_{i,k-i},\qquad k=0,\hdots,r-1,
\end{equation}
\begin{equation}\label{m2}
u_k={1\over 2}\sum_{i=0}^{n-k}{\cal{W}}_{k+i,n-i},\qquad
\alpha_k={1\over 2}\sum_{i=0}^{n-k}\Omega_{k+i,n-i},\qquad k=r,\hdots,n,
%\end{aligned}
\end{equation}
where 
\begin{align*}
{\cal{W}}_{i,j}&=(D\theta_i)+(D\theta_j)+\gamma_i w_{j,x}+
\gamma_j w_{i,x}+\theta_i(Dw_j)+\theta_j(Dw_i),\\
\Omega_{i,j}&=(Dw_i)(D\theta_j)+(Dw_j)(D\theta_i)+\gamma_i (Dw_j)_x+
\gamma_j (Dw_i)_x.
\end{align*}
To have the reduction to the KdV case, we specify
\[
u_n=-1,\qquad \alpha_n=0, 
\]
and we only need to choose $\theta_n=-\vartheta, w_n=0$
for consistence.
Thus, the formulae (\ref{m2}) become
\begin{equation}\label{m3}
\begin{aligned}
u_k&=-1+(D\theta_k)+\eta_n w_{k,x}+{1\over 
2}\sum_{i=1}^{n-k-1}{\cal{W}}_{k+i,n-i},\\
\alpha_k&=-(Dw_k)+\eta_n(Dw_k)_x+{1\over 2}\sum_{i=1}^{n-k-1}\Omega_{k+i,n-i}.
\end{aligned}
\end{equation}

Having the maps constructed, we now obtain
\begin{pro}
Solving the equations (\ref{par}) for $\kappa_i$, the operators
$B_k$(\ref{ha}) is related to a constant coefficient Hamiltonian
operator in 
\[
B_k=M_k{\hat B}_k(M_k)^{\dag},
\]
where $M_k$ is the Fr\'echet derivative of (\ref{m1})(\ref{m3}) and
\[
{\hat B}_k=\left(\begin{array}{cc}
\begin{array}{ccc}0&&-K\\
&\adots&\\ -K&&0\end{array} &0 \\ 0 & \begin{array}{ccc}0&&K\\
&\adots&\\ K&&0\end{array}\end{array}\right),
\]
where $\hat{B}_k$ has the same block structures as $B_k$(\ref{ha}).
\end{pro}
{\em Proof}: Direct computation.

{\em Remarks}:
\begin{itemize}
\item The Hamiltonian nature of our operators $B_k$
is proved as a simple corollary of above proposition for
the generic case. The general case can be proved
as taking limits as in \cite{af3}.
\item The general Miura map (\ref{m1})(\ref{m3}) can be
regarded as decomposition of $n$ step elementary maps as in
\cite{af3}. 
In this way, the remarkable picture of
 \cite{af3}(cf. Fig 1 of \cite{af3})
reappears here.
\end{itemize}

\section{Examples}
In this section, we prsent some interesting examples.
We will concentrate on the simplest cases, that is $n=2$ and
$n=4$ case.

\paragraph{Two Component Case}

In this case, we take $\varepsilon_0=1$, $\varepsilon_1=0$ and
$u(x,t;\lambda) = u(x,t)-\lambda$ and 
$\alpha(x,t;\lambda) = \alpha(x,t)$. Then, we seek the 
formal solution $\eta=\sum_{i=1}^{\infty}\eta_i\lambda^{-i}$ of
the equation
\[
\eta_x+\eta(D\eta)+u\eta-\lambda\eta+\alpha=0,
\]
the first a few solutions, denoted as, are
\begin{align*}
{\cal H}_1&=\alpha,\qquad {\cal H}_2=u\alpha, \\
{\cal H}_3&=u\alpha_x+\alpha(D\alpha+u^2),
\end{align*}
which serve as the first Hamiltonians.
The corresponding first non trivial system($t_2$ flow) is
\begin{equation}\label{4.1}
\begin{aligned}
u_t&=-u_{xx}+2uu_x+2(D\alpha)_x,\\
\alpha_t&=\alpha_{xx}+2(u\alpha)_x,
\end{aligned}
\end{equation}
we note that the above system reduces
to Burgers equation when $\alpha=0$, so it can be regarded as
a supersymmetric Burgers equation.
We also remark that the next flow ($t_3$-flow)
can be transformed to one of $N=2$ supersymmetric KdV equation\cite{lam}
by a invertable change of coordinates\cite{liu3}
 
The Miura map in the present case is the basic
one (\ref{basicmiura}) and the modified system for (\ref{4.1}) is
\[
\left(\begin{array}{cc} w \\ \theta\end{array}\right)_t
=\left(\begin{array}{cc} 0&1\\ -1&0\end{array}\right)
\left(\begin{array}{cc} \delta_w\\ \delta_{\theta}\end{array}
\right)
{\hat{\cal{H}}}_2,
\]
where ${\hat{\cal{H}}}_2=w_x(D\theta)(Dw)
+(D\theta)^2(Dw)+(Dw)(Dw)_x\theta+(D\theta)(Dw)_x$.

To have new example, we choose $\varepsilon$ as before and $u= u_0+\lambda u_1$
and $\alpha =\lambda\alpha_1$ with $u_0$
is a constant. The Hamiltonian operators are
\[
J_0=\left(\begin{array}{cc}2D\partial &-\partial^2+u_0\partial\\
\partial^2+u_0\partial &0\end{array}\right),\qquad
J_1=-\left(\begin{array}{cc}2\alpha_1-(Du_1)&-\alpha_1 D+\partial u_1\\
u_1\partial+D\alpha_1 &\alpha_1 D+D\alpha_1\end{array}\right).
\]

In this case we seek the formal solution
of the form $\eta=\sum_{i=0}^{\infty}\eta_i\lambda^{-i}$
of the equation (\ref{ricati})
and the first two are listed as follows
%\begin{equation}
\begin{align*}
{\cal H}_0&=-{\alpha_1\over u_1},\\
{\cal H}_1&=u^{-1}_{1}\left(
\left({\alpha_1\over u_1}\right)_x-\left({\alpha_1\over u_1}\right)D
\left({\alpha_1\over u_1}\right)+{u_0\alpha_1\over u_1}\right).
\end{align*}
%\end{equation}
With $u_0=c$(constant),
we  have
\begin{align*}
u_{1,t}&=2D\left({\alpha_1\over u_{1}^{2}}\right)_x+\left({1\over 
u_1}\right)_{xx}
-c\left({1\over u_1}\right)_{x},\\
\alpha_{1,t}&=\left({\alpha_1\over u_{1}^{2}}\right)_{xx}+
c\left({\alpha_1\over u_{1}^{2}}\right)_x,
\end{align*}
interestingly, above system admits the reduction $\alpha_1=0$,
which reads as
\begin{equation}\label{b}
v_t=-v^2 (v_{xx}-cv_x),
\end{equation}
with $v=u_{1}^{-1}$. The system(\ref{b}) passes Painleve
test as shown in\cite{cfa}. We also note that when $c=0$, the system(\ref{b})
is the one discussed in \cite{gr} and is in the list of
evolution equations classified in\cite{ss} by symmetry approach.

\paragraph{Four Component Case}
Now we present the last example - system with four component case:
$\varepsilon=1$, $u=u_0+\lambda u_1-\lambda^2$ and $\alpha=
\alpha_0+\lambda\alpha_1$.
Similarly, we have the Hamiltonians
\[
{\cal H}_1=\alpha_1, \qquad {\cal H}_2=\alpha_0+u_1\alpha_1,
\qquad 
{\cal H}_3=u_0\alpha_1+u_{1}^{2}\alpha_{1}+u_1\alpha_0,
\]
and the system are tri-Hamiltonian with
\[
B_0=\left(\begin{array}{cc}J_1&-J_2\\
-J_2&0\end{array}\right),\quad
B_1=\left(\begin{array}{cc}J_0&0\\
0&J_2\end{array}\right),\quad
B_2=\left(\begin{array}{cc}0&J_0\\
J_0&J_1\end{array}\right),
\]
where 
\begin{align*}
J_0&=\left(\begin{array}{cc} 
2D\partial +2\alpha_0-(Du_0)&-\partial^2-\alpha_0 D+\partial u_0\\
\partial^2+u_0\partial +D\alpha_0&\alpha_0\partial +\partial\alpha_0
\end{array}\right),\\
J_1&=\left(\begin{array}{cc} 2\alpha_1-(Du_1)&\alpha_1 D+\partial u_1\\
u_1\partial +D\alpha_1&\alpha_1\partial+\partial\alpha_1\end{array}\right),
\qquad
J_2=\left(\begin{array}{cc} 0&-\partial\\
-\partial&0\end{array}\right),
\end{align*}
the first interesting flow is
\begin{align*}
u_{0,t}&=2(D\alpha_1)_x+2\alpha_0\alpha_1-(Du_0)\alpha_1-u_{1,xx}
-\alpha_0(Du_1)+(u_0u_1)_x,\\
\alpha_{0,t}&=\alpha_{1,xx}+u_0\alpha_{1,x}+D(\alpha_0\alpha_1)
+2\alpha_0u_{1,x}+\alpha_{0,x}u_1,\\
u_{1,t}&=u_{0,x}+2u_1u_{1,x},\\
\alpha_{1,t}&=\alpha_{0,x}+2(u_1\alpha_1)_x.
\end{align*}

The Miura map here reads as
\begin{equation}\label{4.2}
\begin{aligned}
u_0&=w_{0,x}+(D\theta_0)+\theta_0(Dw_0),\\
\alpha_0&=(Dw_0)_x+(D\theta_0)(Dw_0),\\
u_1&=(D\theta_0)+(D\theta_1)-w_{0,x}+w_{1,x}+\theta_0(Dw_1)+\theta_1(Dw_0),\\
\alpha_1&=(Dw_0)(D\theta_1)+(Dw_1)(D\theta_0)-(Dw_0)_x+(Dw_1)_x,
\end{aligned}
\end{equation}
above Miura map(\ref{4.2}) can be decomposed as follows
\begin{align*}
u_0&=v_{0,x}+(D\mu_0)+\mu_0(Dv_0),\\
\alpha_0&=(Dv_0)_x+(D\mu_0)(Dv_0),\\
u_1&=v_1,\\
\alpha_1&=\mu_1,
\end{align*}
and
\begin{align*}
v_0&=w_0,\\
\mu_0&=\theta_0\\
v_1&=(D\theta_0)+(D\theta_1)-w_{0,x}+w_{1,x}+\theta_0(Dw_1)+\theta_1(Dw_0),\\
\mu_1&=(Dw_0)(D\theta_1)+(Dw_1)(D\theta_0)-(Dw_0)_x+(Dw_1)_x.
\end{align*}
Thus, we have two step modifications here.
The modified systems under all these
Miura maps can be easily calculated and we will
not present them here. 

{\em Remark}: The Miura map(\ref{4.2}) is resulted
from the general construction of the section 3.
It is possible to rederive it by linearization
of the basic one. Indeed, linearizing
the basic map(\ref{basicmiura}), we have
\begin{align*}
u_0&={\hat w}_{0,x}+(D{\hat \theta}_0)+{\hat\theta}_0(D{\hat w}_0),\\
\alpha_0&=(D{\hat w}_0)_x+(D{\hat \theta}_0)(D{\hat w}_0),\\
u_1&={\hat w}_{1,x}+(D{\hat{\theta}}_1)+{\hat \theta}_0(D{\hat 
w}_1)+{\hat{\theta}}_1(D{\hat w}_0),\\
\alpha_1&=(D{\hat w}_1)_x+(D{\hat{\theta_1}})(D{\hat w}_0)+(D{\hat 
\theta}_0)(D{\hat w}_1),
\end{align*}
the above map is equivalent to (\ref{4.2}) by a simple
transformation, namely, 
\[
{\hat w}_0=w_0, \quad {\hat\theta}_0=
\theta_0, \quad \hat{w}_1=w_1-w_0,
\quad {\hat{\theta}}_1=\theta_0+\theta_1.
\]
%\newpage

\end{document}